\documentclass[aps,prl,reprint,superscriptaddress]{revtex4-2}
\usepackage{graphicx}
\usepackage{dcolumn}
\usepackage{bm}
\usepackage{color}
\usepackage{amsmath}

\usepackage[mathlines]{lineno}

\begin{document}

\preprint{AAPM/123-QED}

\title{Phonon-driven decoherence of high-harmonic generation in the solid-state}

\begin{abstract}
High-harmonic generation in solids has emerged as a powerful probe of ultrafast electron dynamics and lattice motion, and recent theoretical work has suggested that thermally driven lattice fluctuations can act as an effective source of decoherence in the harmonic-generation process. However, a direct experimental link between high-harmonic emission and temperature-driven incoherent phonons has remained unclear. Here, we investigate the temperature dependence of high-harmonic generation in ultrapure silicon using reflection-geometry measurements over a wide temperature range. We observe that the harmonic yield increases significantly with decreasing temperature. To interpret these results, we introduce a one-dimensional atomic-chain model in which finite temperature is represented by random lattice displacements that mimic incoherent phonon fluctuations. The simulations reproduce the magnitude of temperature-dependent change of the harmonic signal and support a picture in which thermally induced lattice disorder enhances electron-hole decoherence, thereby reducing high-harmonic emission. Our results establish incoherent phonons as an important source of decoherence in solid-state high-harmonic generation.
\end{abstract}
\author{Saadat~Mokhtari}
\affiliation{Joint Attosecond Science Laboratory, National Research Council of Canada and University of Ottawa, 100 Sussex Drive, Ottawa, Ontario K1A 0R6, Canada}
\affiliation{Advanced Laser Light Source (ALLS), Institut National de la Recherche Scientifique, Centre \'{E}nergie Mat\'{e}riaux T\'{e}l\'{e}communications, 1650 Boulevard Lionel-Boulet, Varennes, Qu\'{e}bec J3X 1P7, Canada}

\author{Vedran~Jelic}
\affiliation{Joint Attosecond Science Laboratory, National Research Council of Canada and University of Ottawa, 100 Sussex Drive, Ottawa, Ontario K1A 0R6, Canada}
\affiliation{Advanced Laser Light Source (ALLS), Institut National de la Recherche Scientifique, Centre \'{E}nergie Mat\'{e}riaux T\'{e}l\'{e}communications, 1650 Boulevard Lionel-Boulet, Varennes, Qu\'{e}bec J3X 1P7, Canada}

\author{David~N.~Purschke}
\affiliation{Joint Attosecond Science Laboratory, National Research Council of Canada and University of Ottawa, 100 Sussex Drive, Ottawa, Ontario K1A 0R6, Canada}
\author{Shima~Gholam~Mirzaeimoghadar}
\affiliation{Joint Attosecond Science Laboratory, National Research Council of Canada and University of Ottawa, 100 Sussex Drive, Ottawa, Ontario K1A 0R6, Canada}

\author{Kasia~Kowalczyk}
\affiliation{Joint Attosecond Science Laboratory, National Research Council of Canada and University of Ottawa, 100 Sussex Drive, Ottawa, Ontario K1A 0R6, Canada}

\author{David~A.~Reis}
\affiliation{Department of Applied Physics, Stanford University, Stanford, California 94305, USA}

\author{T.~J.~Hammond}
\affiliation{Department of Physics, University of Windsor, Windsor ON N9B 3P4, Canada}

\author{David~Villeneuve}
\affiliation{Joint Attosecond Science Laboratory, National Research Council of Canada and University of Ottawa, 100 Sussex Drive, Ottawa, Ontario K1A 0R6, Canada}

\author{Andr\'e~Staudte}
\affiliation{Joint Attosecond Science Laboratory, National Research Council of Canada and University of Ottawa, 100 Sussex Drive, Ottawa, Ontario K1A 0R6, Canada}

\author{Fran\c{c}ois~L\'egar\'e}
\affiliation{Advanced Laser Light Source (ALLS), Institut National de la Recherche Scientifique, Centre \'{E}nergie Mat\'{e}riaux T\'{e}l\'{e}communications, 1650 Boulevard Lionel-Boulet, Varennes, Qu\'{e}bec J3X 1P7, Canada}

\author{Giulio~Vampa}
\email{Giulio.Vampa@uottawa.ca}
\affiliation{Joint Attosecond Science Laboratory, National Research Council of Canada and University of Ottawa, 100 Sussex Drive, Ottawa, Ontario K1A 0R6, Canada}
\maketitle

High-harmonic generation is an extremely nonlinear optical process that up-converts the frequency of an intense femtosecond laser to its high order harmonics during interaction with a medium. Although first demonstrated in metals \cite{burnett1977harmonic}, the most famous occurrence was carried out with noble gas atoms \cite{ferray1988multiple}. Gas-phase emission of harmonics was famously interpreted as a three-step process that starts with nonresonant laser-induced ionization of the atoms, followed by acceleration of the ionized electrons to very high kinetic energies and concludes with their recollision and recombination with the parent ions and emission of high-harmonic photons \cite{corkum1993plasma}. The process has since been extended to crystals \cite{Chin2001, ghimire2011observation} and liquids \cite{luu2018extreme}. The discovery of high-harmonic generation in bulk crystals established a new way to investigate ultrafast phenomena in solid-state materials on femtosecond-to-attosecond timescales \cite{ghimire2019high, Ortmann2021, goulielmakis2022high}.
Like in the gas phase, electron--hole pairs are created in the laser field across the material's energy gap and accelerated to high momenta that can even exceed the Brillouin zone, setting up nonlinear intraband and interband currents that radiate high harmonics \cite{vampa2014theoretical}. During their motion, the electron--hole pairs probe the material's electronic structure and symmetry and such information can ultimately be retrieved from the high-harmonic spectrum. High-harmonic spectroscopy in the solid state \cite{Heide2024} has successfully retrieved the timing of the recolliding electrons and holes \cite{vampa2015linking, freudenstein2022attosecond}, the material's band structure \cite{vampa2015all, Lanin2017, lv2021high, uzan2022observation, Parks2025}, Berry curvature \cite{uzan2024observation, luu2018measurement}, and topological \cite{heide2022probing} and structural phase transitions \cite{bionta2021tracking}.

The influence of ionic motion on high-harmonic generation has been extensively studied in the gas phase, where nuclear wavepacket dynamics during molecular vibrations and dissociation modulate both the amplitude and phase of the emitted harmonics, as demonstrated in excited Br$_2$ \cite{worner2010following, worner2010high} and in isotopic mixtures of H$_2/$D$_2$ \cite{baker2006probing}. In solids, meV energies are sufficient to excite collective vibrations of the ions -- phonons. Phonon excitation can either be coherent or incoherent (thermal), both leading to disorder in an otherwise periodic lattice. Recent theoretical \cite{rana2022high, Neufeld2022, hu2024phonon} and experimental \cite{bionta2021tracking, zhang2024high, zhang2024optica, zhang2025acs} work has demonstrated that high-harmonic generation can serve as a sensitive probe of coherent lattice dynamics whereby the high-harmonic spectrum is modulated via phonon-driven changes in the electronic structure.

Since ionic displacements induce changes in the phase of the high harmonics, the disorder associated with incoherent phonons can lead to loss of coherence of the harmonics, i.e. dephasing of the electron--hole polarization arising from electron--phonon scattering. Theoretical studies have examined thermally driven atomic motion as an effective source of electronic decoherence. Freeman et al. \cite{freeman2022high} simulated high-harmonic generation in crystalline silicon and diamond within time-dependent density functional theory and introduced thermal effects by displacing atomic positions in a large silicon supercell using configurations taken from molecular-dynamics simulations. They found that thermally induced atomic motion acts as an effective dephasing mechanism and suggested cooling as a route to enhance high-order harmonic yield. Similarly, Du and Ma \cite{du2022temperature} developed a temperature-dependent lattice-vibration model for solid-state high-harmonic generation and showed that increasing temperature enhances dephasing, effectively increasing electronic decoherence. Recent theoretical work have further supported a picture in which lattice fluctuations act as a source of electronic decoherence in solid-state high-harmonic generation, thereby suppressing the harmonic emission even at low temperature \cite{cardenas2025effects}.

To date, experiments have not established a clear link between harmonic emission and temperature-driven incoherent phonons. However, temperature-dependent high-harmonic generation has been explored in materials undergoing phase transitions, such as superconductors \cite{alcala2022high} and Mott insulators \cite{murakami2022anomalous}. Meng et al. \cite{meng2023higher} studied high-harmonic generation in boron-doped silicon at cryogenic and room temperatures. They observed brighter harmonics at 4 K than at room temperature and attributed the enhanced harmonic yield to a change in the dominant generation mechanism: at high temperature, the response is governed by the nonlinear intraband motion of thermally ionized carriers, whereas at low temperature, dopant tunnel ionization injects carriers into the valence bands, and the harmonic emission arises from both this generational current and the subsequent acceleration of free carriers. \\ 

In this work, we address this experimental gap by measuring the temperature dependence of high-harmonic generation in undoped silicon, a  semiconductor crystal that does not undergo a phase transition upon cooling in the temperature range considered here. To minimize propagation effects in the bulk, the experiment is performed in reflection geometry since the intense driving field becomes easily modified as it propagates through the solid, leading to coherent interference of high-harmonic emission from different depths that strongly reshapes the observed spectrum \cite{floss2018ab}. To interpret our measurements, we introduce a one-dimensional atomic-chain model in which finite temperature enters through random lattice displacements that mimic incoherent phonon fluctuations, and that is in qualitative agreement with the experimental data.

\begin{figure}[!t]
  \begin{flushright}
    \includegraphics[width=\columnwidth]{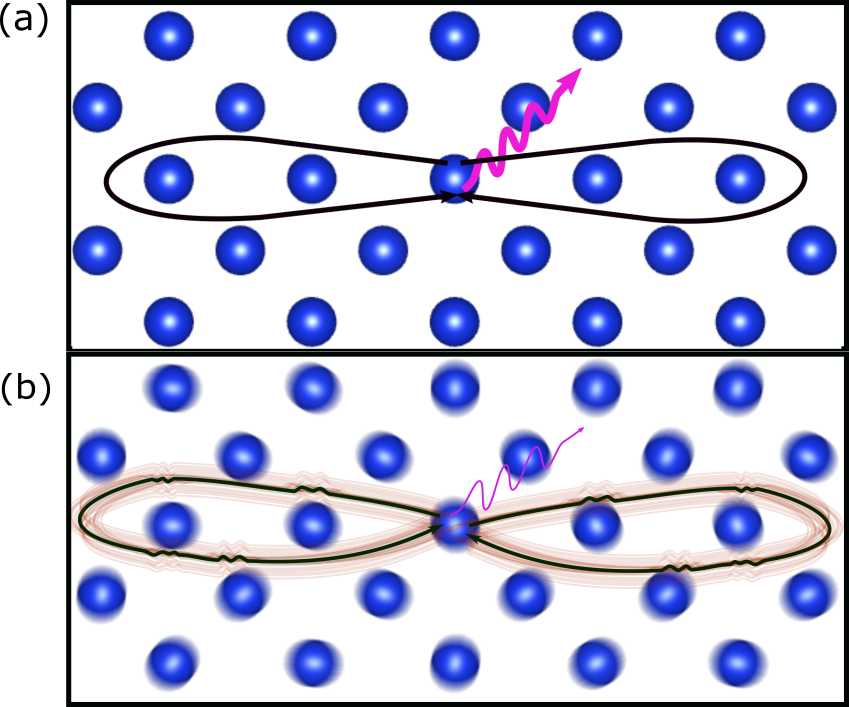}
  \end{flushright}
  \caption{Schematic electron--hole real-space trajectory in silicon at (a) 0~K and (b) 300~K.At 0 K, the lattice is static, and the electron–hole pair follows a well-defined, coherent trajectory (black line), leading to efficient recombination and high-harmonic emission (magenta). At 300 K, thermally activated lattice vibrations introduce fluctuations in the ion positions. The diffuse red line represents the spread of real-space trajectories arising from phonon-induced scattering, reflecting a loss of coherence for the electron--hole pair and suppressing the high-harmonic yield.}
  \label{fig:Schematic}
\end{figure}

Figure~\ref{fig:Schematic} provides a schematic interpretation of how temperature suppresses solid-state high-harmonic generation through thermally induced lattice motion. At 0 K (panel \textbf{a}), the lattice is static, so an electron–hole pair driven by the optical field follows a well-defined excursion in real space (black trajectory). In interband high-harmonic generation, when the electron and hole  recombine, a  high-harmonic photon is emitted \cite{vampa2015linking}. At room temperature (300 K, panel \textbf{b}), incoherent lattice vibrations (phonons) introduce fluctuations of the ions positions that electrons and holes experience as static disorder during the brief quiver motion in the laser field. Disorder-induced scattering causes loss of coherence of the electron-hole pair (represented as blurred lines), which in turn reduces the probability of phase-matched recombination and leads to a weaker high harmonic generation signal at elevated temperature.\\

In the experiment, a Yb:KGW laser (LightConversion Carbide) delivers 260-fs pulses at a center wavelength of 1030 nm with a pulse energy of 0.8 mJ and a repetition rate of 100 kHz. Approximately 75\% of the output pumps an optical parametric amplifier (LightConversion Orpheus-MIR) generating broadband mid-infrared (MIR) pulses centered at 3.2~$\mu$m with a duration of $\sim$60 fs. The MIR beam is focused onto the silicon sample using a gold-coated $90^\circ$ off-axis parabolic mirror ($f$ = 10~cm) at an incidence angle of $\sim10^\circ$. The sample is a 40-$\mu$m-thick high purity silicon crystal with a (100) surface orientation (NORCADA). It is mounted in a liquid nitrogen cryostat (RC102-CFM Microscopy Cryostat) equipped with a calcium fluoride window, allowing temperature control from 77 K to 500 K via PID-regulated resistive heating. The emitted high harmonics are collected in reflection geometry using a UV-enhanced aluminum mirror. The spectrum is analyzed with a home-built spectrometer equipped with a 300 $l$/mm vacuum ultraviolet grating (Acton), an image intensifier (Photonis), and a CCD camera. The MIR pulse energy and polarization are controlled using half-wave plates and wire grid polarizers.

\begin{figure}[!t]
  \begin{flushright}
    \includegraphics[width=\columnwidth]{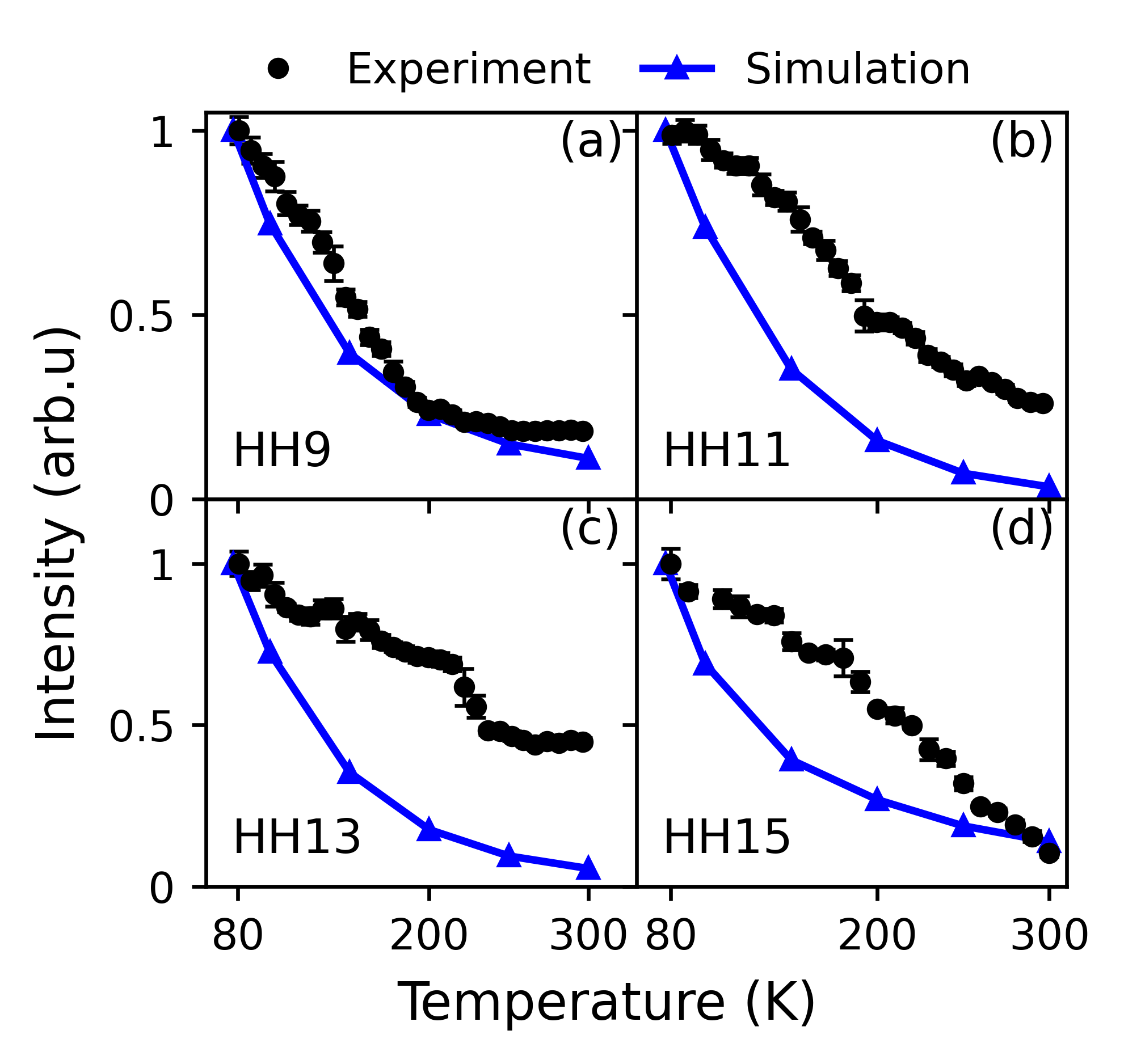}
  \end{flushright}
  \caption{Temperature dependence of the harmonic yield for the 9$^{\text{th}}$ (a), 11$^{\text{th}}$ (b), 13$^{\text{th}}$ (c), and 15$^{\text{th}}$ (d) harmonics. Black circles represent experimental data, and blue triangles show the simulation results. Error bars denote the standard deviation from three independent measurements.}
  \label{fig:Tscan}
\end{figure}

Figure~\ref{fig:Tscan}a--d shows the yield of harmonics 9, 11, 13 and 15 as a function of temperature, normalized to their values at 77 K. The peak incident vacuum  intensity of the MIR pulse is $\sim$0.3~$\mathrm{TW/cm^2}$. The yield of all measured harmonics increases significantly with decreasing temperature. To test our hypothesis that increased disorder at elevated temperatures leads to a loss of coherence of the laser-driven electron--hole pairs, we turn to simulations.\\ 

\begin{figure}[!t]
  \begin{flushright}
    \includegraphics[width=\columnwidth]{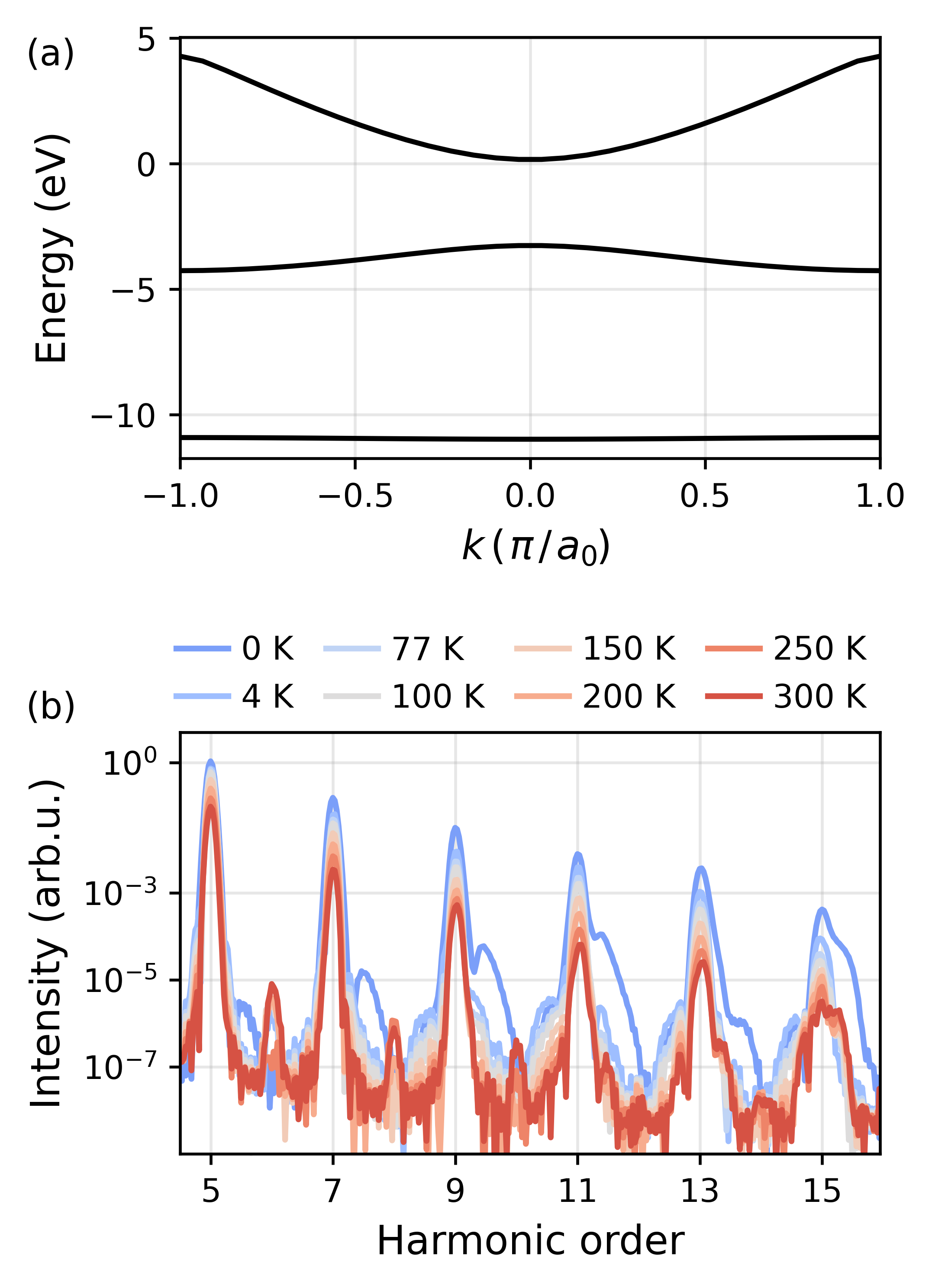}
  \end{flushright}
  \caption{Calculated (a) band structure and (b) high harmonic spectra for different temperatures.}
  \label{fig:simulation}
\end{figure}

To model the impact of lattice temperature on solid-state high-harmonic generation, we numerically solve the time-dependent density matrix of a one-dimensional solid dressed with the laser field. We consider one electron in a one-dimensional lattice potential described by the field-free Hamiltonian
\begin{equation}
\hat{H}_{0}=\frac{\hat{p}^{2}}{2m}+V(x),
\label{eq:H0}
\end{equation}
with
\begin{align}
&V(x)=-V_{0}\sum_j\left[1+\cos\!\left(\frac{2\pi}{a_{0}}(x-ja_{0})\right)\right],
\label{eq:Vx} \\
& f(x) = 
\begin{cases}
    1+\cos\!\left[\frac{2\pi}{a_{0}}(x-ja_{0})\right] &\text{if } |x| \leq ja_0 + a_0/2 \\
    0 &\text{if } |x| >  ja_0 + a_0/2
\end{cases}
\end{align}
where the lattice constant $a_{0}=5.43~\mathrm{\AA}$, and $V_{0}=7.5$~eV is chosen to yield a band gap and band widths close to those of silicon. The lattice potential is modeled as a sum of single-site cosine wells, each truncated to one unit cell centered on the corresponding ion. Given this potential, the band structure and eigenstates are obtained by diagonalizing the field-free Hamiltonian. Figure~\ref{fig:simulation}a shows the calculated band structure. It features three bands and a direct band gap of 3.4 eV. The transition dipole moments between the bands are calculated from the eigenstates. Finite temperatures populate phonons with random phases, leading to incoherent lattice fluctuations. We capture these ions displacements by shifting the potential minima $ja_{0}\to ja_{0}+\Delta x_j$. The random displacement amplitude $\Delta x_j$ is drawn from a Gaussian distribution whose standard deviation is $u_{\mathrm{rms}}(T)=\sqrt{\langle u^{2}\rangle}$, where the mean-squared displacement $\langle u^{2}\rangle$ is that of a quantum harmonic oscillator:
\begin{equation}
\langle u^{2}\rangle=\frac{3}{m\hbar\omega}\left[n(T)+\frac{1}{2}\right],
\label{eq:u2}
\end{equation}
where $m = 28.08$ amu is the silicon atomic mass, $\hbar\omega = 3.04 \times 10^{-4}$ Ha is the phonon energy, and $n(T)$ is the Bose-Einstein occupation at temperature $T$ \cite{kittel2018introduction}. For reference, the largest $\langle u\rangle$ = $0.41\,\mathrm{\AA}$  at 300 K.
The laser field is incorporated in the velocity gauge via minimal coupling, yielding the time-dependent Hamiltonian $\hat H(k,t)=\hat H_0(k)-A(t)\hat p(k)$, where $A(t)=-\int_{-\infty}^{t}F(t')\,dt'$ is the laser vector potential. The system is driven by an electric field $F(t)$ with central wavelength $\lambda=3.2~\mu\mathrm{m}$ and peak amplitude $F_{0}=0.003$~a.u. The density matrix $\rho_k(t)$ is propagated according to the Liouville--von Neumann equation,
$\dot{\rho}_k(t)=-i[\hat{H}(k,t),\rho_k(t)]$,
which is numerically integrated using a fourth-order Runge--Kutta (RK4) scheme \cite{orlando2020simple,du2021high}, with the initial condition $\rho_k(0)$ corresponding to fully occupied valence bands and empty conduction band. The time-dependent current density is obtained from the momentum expectation value and averaged over the Brillouin zone,
\begin{equation}
j(t)=\mathrm\,\big\langle \mathrm{Tr}\!\big[\rho_{k}(t)\,\hat{p}(k)\big]\big\rangle_{k}.
\label{eq:current}
\end{equation}

Because a single thermally displaced configuration generally breaks inversion symmetry and can generate strong even harmonics, we perform a coherent ensemble average of the time-domain current over 1000 independent configurations, $\overline{J}(t)=\langle j(t)\rangle_{\mathrm{seeds}}$. This configuration average statistically restores centrosymmetry and strongly suppresses even-order emission (no even harmonics are observed in the experiment). The high-harmonic spectrum is finally obtained from the squared modulus of the Fourier transform, $|\mathcal{F}\{\overline{J}(t)\}|^{2}$. Spectra for various temperatures are reported in Fig. \ref{fig:simulation}b. The power of each harmonic is integrated over the spectral width, and the results are overlayed to the experimental data in Fig. \ref{fig:Tscan}, blue triangles. The simulations capture the global trend of decreasing high-harmonic intensity with increasing temperature. Although our current experimental capability limits the lowest accessible temperature to 77 K, our model predicts a 15–30-fold enhancement across the high-harmonic spectrum at liquid-helium temperature.
The discrepancy between simulation and experiment is expected since the the model employed here is a one dimensional three-band model (Figure~\ref{fig:simulation}a) where the emitted spectrum is computed from interband dynamics only. Despite these simplifications, the agreement in the overall temperature scaling supports thermal decoherence as a dominant mechanism for the observed intensity suppression of solid-state high-harmonic generation.\\

In conclusion, we have experimentally demonstrated a clear temperature dependence of solid-state high harmonic generation in ultrapure silicon, with the power of harmonics increasing significantly as the temperature is decreased from 300 K to 77 K. The remarkable quantitative agreement with a theoretical modeling based on a one-dimensional lattice with random displacements supports the hypothesis that thermally populated incoherent phonons are the microscopic origin of this suppression. In this picture, increasing temperature enhances lattice fluctuations that act as static disorder during the sub-cycle electron-hole motion, thereby reducing electron-hole phase coherence, lowering the probability of coherent recombination and resulting in weaker high-harmonic emission. These results establish high-harmonic generation as a sensitive probe of incoherent-phonon-driven decoherence in a conventional semiconductor and provide a route to study how lattice disorder limits carrier coherence and high harmonic emission in solids.

\bibliographystyle{apsrev4-2}
\bibliography{refs}

\end{document}